\documentclass{aa}
\usepackage{natbib}
\usepackage{graphicx}
\bibpunct{(}{)}{;}{a}{}{,} 

\begin{document}

\title{The disk-bearing young star IM Lup \thanks{Based on observations obtained with XMM-Newton, an ESA science mission, and Chandra, a NASA science mission, both with instruments and contributions directly funded by ESA Member States and NASA .}}
\subtitle{X-ray properties and limits on accretion}

\author{H.~M. G\"unther \inst{1,2}  \and S.~P.~Matt \inst{3} \and J.~H.~M.~M.~Schmitt \inst{1} \and M.~G\"udel \inst{4,5} \and Z.-Y.~Li \inst{6} \and D.~M.~Burton \inst{7}}  
\offprints{H.~M. G\"unther,\\ \email{moritz.guenther@hs.uni-hamburg.de}}
\institute{Hamburger Sternwarte, Universit\"at Hamburg, Gojenbergsweg 112, 21029 Hamburg, Germany \and 
Harvard-Smithsonian Center for Astrophysics, 60 Garden Street, Cambridge, MA 02138, USA \and
NASA Ames Research Center, Moffett Field, CA 94035, USA \and
Institute of Astronomy, ETH Z\"urich, 8093 Z\"urich, Switzerland \and 
Department of Astronomy, University of Vienna, T\"urkenschanzstr. 17, 1180 Vienna, Austria \and 
Department of Astronomy, University of Virginia, P.O. Box 400325, Charlottesville, VA 22904, USA \and
University of Southern Queensland, Toowoomba Qld 4350, Australia} 
\date{Received / Accepted }
\abstract{Classical T Tauri stars (CTTS) differ in their X-ray signatures from older pre-main sequence stars, e.g. weak-lined TTS (WTTS). CTTS show a soft excess and deviations from the low-density coronal limit in the He-like triplets.}{We test whether these features correlate with accretion or the presence of a disk by observing IM~Lup, a disk-bearing object apparently in transition between CTTS and WTTS without obvious accretion.}{We analyse a \emph{Chandra} grating spectrum and additional \emph{XMM-Newton} data of IM~Lup and accompanying optical spectra, some of them taken simultaneously to the X-ray observations. We fit the X-ray emission lines and decompose the H$\alpha$ emission line in different components.}{In X-rays IM~Lup has a bright and hot active corona, where elements of low first-ionisation potential are depleted. The He-like \ion{Ne}{ix} triplet is in the low-density state, but due to the small number of counts a high-density scenario cannot be excluded on the 90\% confidence level. In all X-ray properties IM~Lup resembles a  main-sequence star, but it is also compatible with CTTS signatures on the 90\% confidence level, thus we cannot decide if the soft excess and deviations from the low-density coronal limit in the He-like triplets in CTTS require accretion or only the presence of a disk. IM~Lup is chromospherically active, which explains most of the emission in H$\alpha$. Despite its low equivalent width, the complexity of the H$\alpha$ line profile is reminiscent of CTTS. We present an estimate for the mass accretion rate of $10^{-11}M_{\sun}$~yr$^{-1}$. }{}

\maketitle

\section{Introduction}
\label{introduction}
Our view on the formation of stars and planetary systems and their emergence from molecular clouds has made significant progress over the last decades. Stars form when giant molecular clouds fragment and contract to form proto-stars. Mass accretion onto those stellar cores proceeds via an accretion disk, while the surrounding envelope eventually disperses. The low-mass pre-main sequence stars in this stage come in two types, the classical T~Tauri stars (CTTS) and the weak-lined T~Tauri stars (WTTS). Traditionally, they were distinguished only by their H$\alpha$ emission line equivalent width, with those stars of EW $> 10$~\AA{} defined to be CTTS. It turned out that the EW of H$\alpha$ is a good tracer of the accretion flow from the circum-stellar disk the CTTS still possess. The H$\alpha$ EW of WTTS is smaller and their line profiles are symmetric; in contrast CTTS exhibit broader emission lines, which are sometimes asymmetric \citep{2009A&A...504..461F}. Usually the combination of strong H$\alpha$ emission and line asymmetry is a reliable accretion indicator \citep{1998ApJ...492..743M,2003ApJ...592..266M}.

WTTS with their low H$\alpha$ EW are generally expected not to show accretion. It is suggestive to interpret the WTTS as more evolved CTTS, where accretion has stopped already. This does not necessarily imply the absence of a disk, as transitional disks may exist without ongoing accretion for some time \citep{2006ApJ...645.1283P}. For many TTS the H$\alpha$ EW is known to be variable, possibly because the accretion switches on and off or --at least-- the accretion rate changes. 

Both types of TTS have been known for a long time to be copious X-ray emitters \citep{1999ARA&A..37..363F}. As a class CTTS stand out from other X-ray sources by their strong soft X-ray excess \citep{RULup,manuelnh}, but CTTS also exhibit normal coronal activity and stellar flares.

The CTTS TW~Hya, observed with \emph{Chandra}/HETGS and \emph{XMM-Newton} \citep{2002ApJ...567..434K,twhya}, was the first star where unusual line ratios in the He-like triplets was found indicating high densities in the formation region. This phenomenon repeats in most CTTS observed so far (\object{BP Tau}: \citet{bptau}; \object{V4046 Sge}: \citet{v4046}; \object{RU Lup}: \citet{RULup}; \object{MP Mus}: \citet{2007A&A...465L...5A}; \object{Hen 3-600}: \citet{2007ApJ...671..592H}).
There are exceptions to this rule -- in the more massive, eponymous T~Tau itself, although known for its high accretion rate, the \ion{O}{vii} triplet is consistent with the coronal limit \citep{ttau}. The same has been found for HAeBe stars, which are in a similar evolutionary state as the CTTS, yet of higher mass (\object{AB Aur}: \citet{ABAur}; \object{HD 163296}: \citet{HD163296}). The soft excess and the He-like triplet line ratios in CTTS can be naturally linked to mass accretion from proto-planetary disks \citep{lamzin,acc_model}.

Systems containing typical WTTS such as \object{TWA 4} \citep{2004ApJ...605L..49K} and \object{TWA 5} \citep{twa5} show no sign of active accretion and no high densities in their X-ray spectra. Well studied examples of young main-sequence (MS) stars such as AU~Mic, Speedy~Mic and AB~Dor \citep{2004A&A...427..667N} also show triplets indicating low densities similar to older MS objects. Still, 5 of 83 WTTS observed by \citet{2006ApJ...645.1283P} do show an IR-excess caused by dust within a few AU from the star, detected by \emph{Spitzer}, thus there is a matter reservoir for accretion. 

Observationally, we could not distinguish in the past if the X-ray signatures observed in CTTS in contrast to WTTS and young MS stars are caused by active accretion or the presence of a disk. To solve this question we observed a system with \emph{Chandra}, that was previously classified as disk-bearing, non-accreting WTTS. Simultaneous to the X-ray observations, the accretion state is observed using optical spectroscopy.

The properties of the our target IM~Lup are summarised in Sect.~\ref{imlupprop}. We then present the X-ray observations of IM~Lup and the accompanying optical data in Sect.~\ref{observations}. We show our new results and compare IM~Lup to CTTS, WTTS and MS stars in Sect.~\ref{results}. In Sect.~\ref{discussion} we discuss the implications and end with a short summary in Sect.~\ref{summary}.

\section{Properties of IM Lup}
\label{imlupprop}
In this article we present new observations of \object{IM Lup} (=Sz~82), the only known X-ray bright transition object with a low H$\alpha$ EW (4~\AA{})\citep{1999MNRAS.307..909W}, far less than the values for typical CTTS such as BP~Tau and TW~Hya. Still, the IR-excess in IM~Lup indicates the presence of a disk \citep{2006ApJ...645.1283P}, making it a favourite object for disk evolution studies in the IR \citep[starting with][]{2006ApJ...645.1283P,2006A&A...456..535S}. 

There is good evidence that IM~Lup is a single object both from ground-based speckle observations \citep{1997ApJ...481..378G} and from space-based \emph{HST} data \citep{2008A&A...489..633P}. \citet{1998MNRAS.301L..39W} derived a \emph{HIPPARCOS} distance of $190\pm27$~pc, larger than earlier estimates, but still consistent to them within the errors. We use this value throughout the paper. The spectral type of the star is M0, however, there is some debate on its mass. The values vary with the pre-main sequence evolutionary tracks adopted: Based on the data of \citet{1994AJ....108.1071H} the models of \citet{1994ApJS...90..467D} yield $M=0.29 M_{\sun}$ and the models of \citet{2000A&A...358..593S} give $M=0.52 M_{\sun}$. \citet{2008A&A...489..633P} use the tracks from \citet{1998A&A...337..403B} and find $M=1 M_{\sun}$ and $R=3 R_{\sun}$. The age is around 1 Myr in all models. Observationally there is support for a mass of $M=1.2\pm0.4 M_{\sun}$ from millimetre observations, which resolve the emission lines from a disk in Keplerian rotation; the inclination of this disk is $54^{\circ}\pm3^{\circ}$ \citep{2009A&A...501..269P}. The reddening has been derived between $A_V=0.5$ \citep{2008A&A...489..633P} and $A_V=1$ \citep{1994AJ....108.1071H}.

The inner disk of IM~Lup is resolved down to $<0.4$~AU, dust grains have partially settled in the disk-midplane and grown to a few millimetres; the disk surface shows strong silicate emission bands and there are signatures of a low degree of crystallisation \citep{2008A&A...489..633P}. Beyond 400~AU the surface density of CO gas and dust grains with sizes 1$\mu$m-1mm drops dramatically. It remains detectable out to 900~AU \citep{2009A&A...501..269P}.

\section{Observations and data reduction}
\label{observations}
\subsection{X-ray observations}
IM~Lup was observed off-axis three times with \emph{XMM-Newton} giving a total exposure time of 84~ks and with \emph{Chandra} for 150~ks in a dedicated observation using the high-energy transmission grating (HETGS). The observation log is summarised in table~\ref{obslog}.
\begin{table}
\caption{\label{obslog}Observing log of X-ray observations.}
\begin{center}
\begin{tabular}{lllr}
\hline \hline
Observatory & ObsID  & Obs. date & Exp. time \\
\hline

XMM-Newton & 0303900301 & 2005-08-08 & 30 ks\\
XMM-Newton & 0303900401 & 2005-08-17 & 27 ks\\
XMM-Newton & 0303900501 & 2005-09-06 & 27 ks\\
Chandra & 9938 & 2009-05-04 & 150 ks\\
\hline
\end{tabular}
\end{center}
\end{table}
All data was reduced using the standard \emph{XMM-Newton} Science Analysis System (SAS) software, version 9.0,  \citep{2004ASPC..314..759G} or the \emph{Chandra} Interactive Analysis of Observations (CIAO) software, version 4.1.2 \citep{2006SPIE.6270E..60F}, in the case of the \emph{Chandra} observation. The \emph{XMM-Newton} observations are contaminated by proton flares and we applied the standard selection criteria. We extracted CCD spectra for the PN and the MOS2 detector. During the observations one of the chips in the MOS1 camera was not operating and, unfortunately, IM~Lup is located in this gap for two of the three observations. We primarily used the MOS2 data for lightcurves because it is less affected by the proton noise and the PN data for spectra because of their higher signal. We extracted lightcurves and CCD spectra from all three observations. We did not reprocess the \emph{Chandra} \texttt{evt2} file, thus the predefined selection criteria apply. Positive and negative orders for the HEG and MEG are merged. Spectral fitting was carried out using XSPEC V12.5.0 \citep{1996ASPC..101...17A}, and individual line fluxes were measured using the CORA line fitting tool \citep{2002AN....323..129N}. CORA employs a maximum likelihood method, taking the Poisson statistics into account. Because the line widths are dominated by instrumental broadening, we keep them fixed at $\Delta\lambda=0.02$~\AA{} for the MEG and $\Delta\lambda=0.01$~\AA{} for the HEG, the line profile is given by a modified Lorentz profile with $\beta=2.5$. We fix the wavelengths at the theoretical values taken from the CHIANTI~5.1 database \citep{CHIANTI,CHIANTIVII}, but by fitting the strongest lines with a free wavelength we verified that our wavelength scale is accurate. The error introduced by fixing the wavelength is negligible compared to the uncertainties from counting statistics in all cases.

\subsection{Optical observations}
\begin{table}
\caption{\label{obslogopt}Log of optical observations.}
\begin{center}
\begin{tabular}{llllr}
\hline \hline
No & Observatory & ID  & Obs. time & Exp. time\\
\hline
H1 & ESO/HARPS & 6201 & 2008-05-09 02:19 & 20 min\\
H2 & ESO/HARPS & 6381 & 2008-05-09 08:55 & 20 min\\
H3 & ESO/HAPRS & 6711 & 2008-05-10 02:18 & 20 min\\
H4 & ESO/HARPS & 6971 & 2008-05-10 08:10 & 20 min\\
A1 & ANU & 35 & 2009-05-05 12:38 & 30 min \\
A2 & ANU & 36 & 2009-05-05 13:09 & 30 min \\
A3 & ANU & 36 & 2009-05-06 11:50 & 30 min \\
A4 & ANU & 37 & 2009-05-06 12:20 & 30 min \\
\hline
\end{tabular}
\end{center}
\end{table}
Observations were carried out using the 79g/mm echelle spectrograph (R$\approx24,000$) on the Nasmyth-B focus of the Australian National University's 2.3m~telescope (ANU) on the nights of May 5th and 6th, 2009. Each night two 1800~s exposures were taken one after another. The observing log for all optical observations is shown in table~\ref{obslogopt}. A GG490 filter is used in the beam to avoid overlapping of the first two orders at the reddest wavelength. A cross disperser grating of 316/750nm was used to allow collection of data of orders 55-34 or 4120-6630~\AA{}.  Bias frames, ThAr~Arc for wavelength calibration and quartz lamp exposures for flat-fielding were also included for the process of data reduction. The code used to reduced the 2.3m data is a slightly modified version of ESpRIT (Echelle Spectra Reduction: an Interactive Tool) \citep{1997MNRAS.291..658D}.

For comparison purposes we also downloaded observations from the ESO 3.6m telescope at La Silla obtained with HARPS for IM~Lup from the ESO archive (program ID 081.C-0779(A)). These datasets are fully reduced. The uncertainty of the flux in these observations was taken as the Poisson error of the count statistic. This is consistent with an estimate from the spectra, where we chose an apparently line-free region close to the H$\alpha$ line. We calculate the mean flux and the standard deviation. The relative error in the spectra is 3\% for each bin in the oversampled spectrum, that is a S/N about 100 rebinned to instrumental resolution, except for the second exposure, where the relative error is 5\%.

To compare IM~Lup to MS stars, we obtained spectra of HD~156274 and HD~156026 from the library of \citet{2003Msngr.114...10B}.

\section{Results}
\label{results}
\subsection{The X-ray lightcurve}
\begin{figure}
\resizebox{\hsize}{!}{\includegraphics{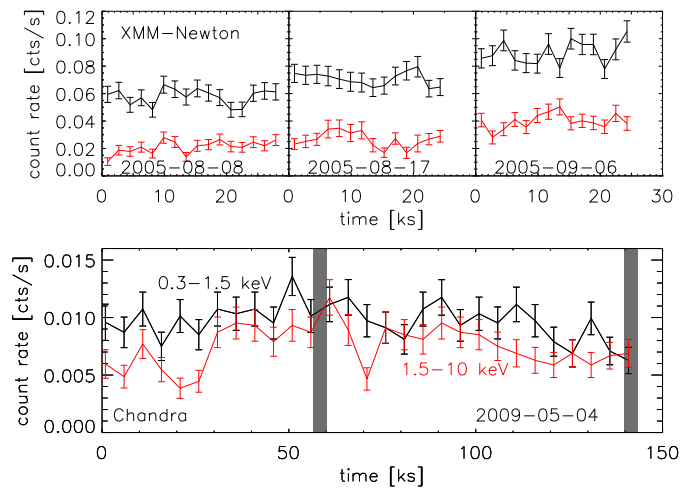}}
\caption{Lightcurve in soft (black) and hard (red/grey) band. Due to different detector efficiencies \emph{XMM-Newton} and \emph{Chandra} cannot be compared directly. The shaded areas show the times of the optical observations.}
\label{fig_lc}
\end{figure}

In Fig.~\ref{fig_lc} the X-ray lightcurve of IM~Lup is shown. The \emph{XMM-Newton} observations are separated by several days and IM~Lup shows a steady increase in luminosity during this time, holding the ratios between hard (1.5-10.0~keV) and soft (0.3-1.5~keV) emission nearly constant. Due to different detector efficiencies the hardness ratios of \emph{Chandra} and \emph{XMM-Newton} cannot be compared directly. During the much longer \emph{Chandra} observation, there is small scale variability in both bands, but without clear flaring activity. A KS-test shows, that a constant lightcurve can be rejected on the 72~\% level for the soft band  and on the 98~\% level for the hard band.

\subsection{The CCD X-ray spectra}
\begin{figure}
\resizebox{\hsize}{!}{\includegraphics[angle=-90]{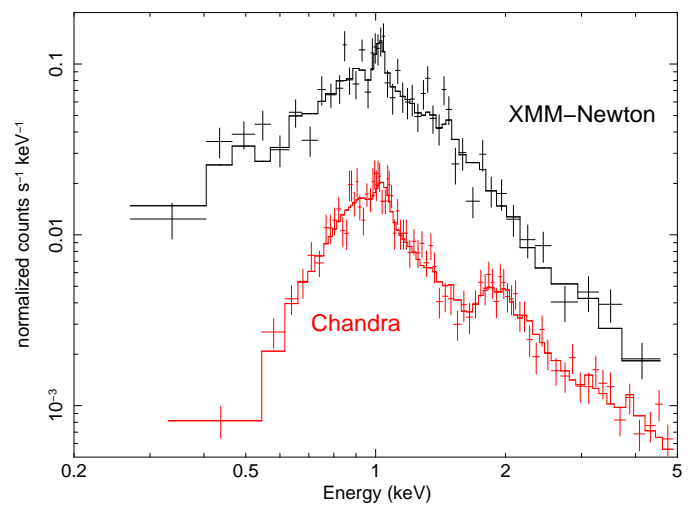}}
\caption{CCD spectra of IM~Lup from \emph{XMM-Newton} EPIC/PN (ObsID 0303900401, upper curve) and \emph{Chandra} ACIS (ObsID 9938, lower curve) with the best-fit model overlaid. See table~\ref{tabfit} for model parameters.}
\label{fig_xspec}
\end{figure}
In Fig.~\ref{fig_xspec} we show the MOS2 CCD spectrum from the second \emph{XMM-Newton} exposure and the zeroth order ACIS spectrum, both binned to a minimum of 25 counts per bin. At first sight we can already identify the \ion{Ne}{x} Ly$\alpha$ line, causing a peak around 1~keV and the \ion{Si}{xiii} and \ion{Si}{xiv} emission around 2~keV. 
We describe the plasma in more detail with three optically thin emission components according to the \texttt{vapec} model combined with a single photoabsorption model in XSPEC and overplot the best-fit model. A simpler model with only two components and coupled absorbing column density $n_{\mathrm{H}}$, temperature, and abundances for all four observations already provides a statistically acceptable fit (red. $\chi^2=1.12$). However, the residuals show systematic trends for low energies indicating an inaccurate combination of $n_{\mathrm{H}}$ and $T_1$. 
Thus we added a third \texttt{vapec} component, keeping the temperatures of the three components coupled over all datasets. This ensures that the relative contribution of soft, medium and hard component can be compared. There is no significant difference in the fitted $n_{\mathrm{H}}$ between the separate datasets, thus we also keep it coupled. The parameters of our best-fit model are given in table~\ref{tabfit}. The errors are calculated as 90~\% confidence ranges and the fitted abundances are relative to the solar values from \citet{1998SSRv...85..161G}; all other abundances are fixed at their respective solar value. Some of the values in the table are strongly coupled. In order to retain the overall luminosity of the star, when e.g. the soft component is lowered the emission measure ($EM$) of the medium component has to be increased. In all observations the emission is dominated by a hot plasma component around k$T=2.1$~keV. The separation between the consecutive \emph{XMM-Newton} exposures is 9 and 19 days. In this time the total plasma emission measure increases by 60\%; at the same time a larger portion of the total emission measure is concentrated at higher temperatures. The lightcurves in Fig.~\ref{fig_lc} show, that this increase is steady during the observations. In the \emph{Chandra} observation the distribution of the emission measure in the different components is comparable to the second \emph{XMM-Newton} dataset, only the hot component is reduced by about a third; thus, the total emission measure is lower than in any of the \emph{XMM-Newton} observations.
\begin{table} [h]
\caption{Best-fit model for the CCD spectra (90\% conf. ranges)\label{tabfit}}
\begin{center}
\begin{tabular}{lrrrr}
\hline\hline
 & \multicolumn{3}{c}{XMM-Newton} & Chandra\\
 ObsID & ...301 & ...401 & ...501 & 9938 \\
\hline
 & \multicolumn{4}{c}{Abundances}\\
\hline
 O  & \multicolumn{4}{c}{$0.5^{+0.3}_{-0.2}$} \\
 Ne & \multicolumn{4}{c}{$2.0^{+0.6}_{-0.5}$} \\
 Mg & \multicolumn{4}{c}{$0.8^{+0.5}_{-0.4}$} \\
 Si & \multicolumn{4}{c}{$0.2^{+0.2}_{-0.2}$} \\
 S  & \multicolumn{4}{c}{$0.1^{+0.2}_{-0.1}$} \\
 Fe & \multicolumn{4}{c}{$0.4^{+0.2}_{-0.1}$} \\
\hline
$n_{\mathrm{H}}$ [$10^{21}$~cm$^{-2}$]& \multicolumn{4}{c}{$1.3\pm0.2$  } \\
k$T_1$ [keV]& \multicolumn{4}{c}{$0.3^{+0.1}_{-0.3}$} \\
k$T_2$ [keV]& \multicolumn{4}{c}{$0.6^{+0.1}_{-0.1}$ }\\
k$T_3$ [keV]& \multicolumn{4}{c}{$2.1^{+0.2}_{-0.1}$} \\
$EM_1$ [$10^{52}$ cm$^{-3}$]& $6^{+9}_{-6}$ & $3^{+13}_{-3}$ & $0^{+35}    $ & $3^{+8}_{-3}$ \\
$EM_2$ [$10^{52}$ cm$^{-3}$]& $5^{+8}_{-5}$ & $10^{+8}_{-10}$ & $12^{+4}_{-4}$ & $10^{+7}_{-4}$ \\
$EM_3$ [$10^{52}$ cm$^{-3}$]& $34^{+6}_{-4}$ & $38^{+6}_{-5}$ & $60^{+7}_{-7}$ & $26^{+4}_{-3}$ \\
\hline
\multicolumn{5}{c}{unabsorbed luminosities (0.3-10 keV)}\\
\hline
$\log L_X$ [erg s$^{-1}$]& 30.7 & 30.8 & 30.9 & 30.7\\
\hline
\multicolumn{5}{c}{$\chi^2=266$  for 254 dof (red. $\chi^2=1.05$)}\\
\hline
\end{tabular}
\end{center}
\end{table}
We measure the absorbing column density towards the source as $N_{\mathrm{H}}=1.3\pm0.2 \times 10^{21}$~cm$^{-3}$. We use this value to calculate the total unabsorbed X-ray luminosities $L_X$ given in table~\ref{tabfit}. Assuming a standard gas-to-dust ratio the optical reddening and the X-ray absorption should be related through the formula
$ N_{\mathrm{H}}=A_V \cdot2\times 10^{21}\mathrm{cm}^{-2}$ \citep{2003A&A...408..581V}, leading to $A_V=0.7$~mag in agreement with the values found by \citet{1994AJ....108.1071H} and \citet{2008A&A...489..633P}.

\subsection{The high-resolution X-ray spectra}
\subsubsection{Line fluxes}
\begin{figure*}
\resizebox{\hsize}{!}{\includegraphics{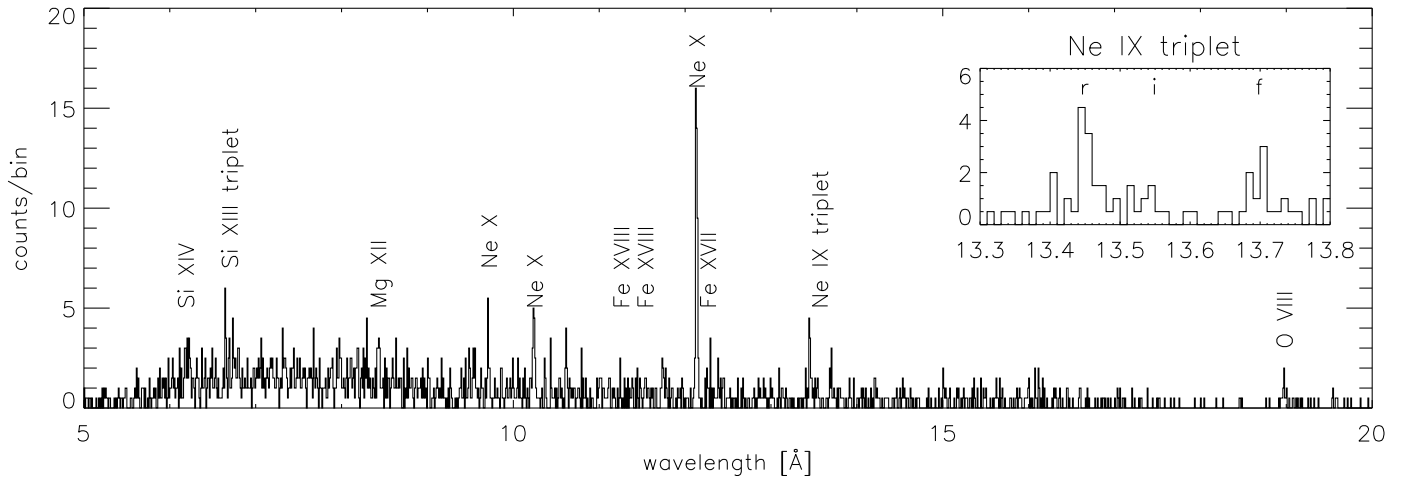}}
\caption{MEG spectrum of IM Lup, rebinned by 2. The lines detected with more than 2$\sigma$ significance are marked (table~\ref{tab1}). The inset shows an enlargement of the He-like \ion{Ne}{ix} triplet.}
\label{fig_meg}
\end{figure*}

The MEG grating spectrum is shown in Fig.~\ref{fig_meg}.
We measure line fluxes from the HEG and the MEG. In table~\ref{tab1} we give fluxes for all lines which were detected above the $2\sigma$ level. In the case of the He-like triplets we quote all lines, if one of them is strong enough to be included in the table. From the fitted line fluxes we calculate the total line luminosity, using the absorption cross section from \citet{1992ApJ...400..699B}. The relative errors on the line intensity are larger than those on the line fluxes, because here the uncertainty in the distance and the fitted $N_{\mathrm{H}}$ contribute to the total error budget.
  \begin{table} [h]                                                                       
  \caption{Measured line fluxes (1$\sigma$ errors)\label{tab1}}                                              
  \begin{center}                                                                          
  \begin{tabular}{lrrrr}                                                                   
  \hline \hline                                                                           
  Ion & $\lambda$ [\AA] & counts & photon flux$^a$ & luminosity$^b$\\                                               
  \hline                                                                                  
  \multicolumn{5}{c}{--- HEG ---}\\                                                       
  \hline
  \ion{Si}{xiv} &  6.18 & $  7.6 \pm   3.3 $ & $     1.2 \pm   0.5 $ & $  17\pm    9$\\                  
  \ion{Mg}{xii} &  8.42 & $  9.0 \pm   3.4 $ & $     1.5 \pm   0.6 $ & $  17\pm    8$\\                  
  \ion{Ne}{x} &  9.71 & $  6.1 \pm   2.9 $   & $   1.8 \pm     0.8 $ & $  19\pm   10$\\                    
  \ion{Ne}{x} & 10.24 & $  5.6 \pm   2.5 $   & $   2.6 \pm     1.2 $ & $  27\pm   14$\\                    
  \ion{Ne}{x} & 12.13 & $ 30.5 \pm   5.7 $   & $  20.3 \pm     3.8 $ & $ 197\pm   68$\\                    
  \hline                                                                                  
  \multicolumn{5}{c}{--- MEG ---}\\
  \hline
  \ion{Si}{xiv} &  6.18 & $ 12.2 \pm   4.7 $ & $     0.9 \pm     0.3 $& $  13\pm    6$  \\                  
  \ion{Si}{xiii} &  6.65 & $ 18.0 \pm   5.2 $& $     1.2 \pm     0.4 $& $  17\pm    7$  \\                 
  \ion{Si}{xiii} &  6.69 & $  7.5 \pm   4.0 $& $     0.5 \pm     0.3 $& $   7\pm    4$  \\                 
  \ion{Si}{xiii} &  6.74 & $ 15.3 \pm   5.1 $& $     0.9 \pm     0.3 $& $  12\pm    5$  \\                 
  \ion{Mg}{xii} &  8.42 & $ 14.9 \pm   5.0 $ & $     0.9 \pm     0.3 $& $  10\pm    4$  \\                  
  \ion{Ne}{x} &  9.71 & $ 14.0 \pm   4.7 $   & $     1.4 \pm     0.5 $& $  14\pm    6$  \\                    
  \ion{Ne}{x} & 10.24 & $ 22.6 \pm   5.7 $   & $     2.6 \pm     0.6 $& $  26\pm   10$  \\                    
  \ion{Fe}{xviii} & 11.25 & $  6.7 \pm 3.0 $ & $     1.1 \pm     0.5 $& $  11\pm    6$  \\                
  \ion{Fe}{xviii} & 11.53 & $  8.9 \pm 3.4 $ & $     1.5 \pm     0.6 $& $  15\pm    7$  \\
  \ion{Ne}{x} & 12.13 & $ 90.8 \pm   9.7 $   & $    19.7 \pm     2.1 $& $ 191\pm   59$  \\
  \ion{Fe}{xvii} & 12.26 & $ 10.2 \pm   3.8$ & $     2.3 \pm     0.9 $& $  22\pm   10$  \\
  \ion{Ne}{ix} & 13.45 & $ 20.9 \pm   5.0 $  & $     7.1 \pm     1.7 $& $  70\pm   26$  \\
  \ion{Ne}{ix} & 13.56 & $  3.3 \pm   2.6 $  & $     1.2 \pm     0.9 $& $  12\pm   10$  \\
  \ion{Ne}{ix} & 13.70 & $ 10.9 \pm   3.9 $  & $     4.0 \pm     1.4 $& $  40\pm   18$  \\
  \ion{O}{viii} & 18.97 & $  7.9 \pm   3.3 $ & $     9.9 \pm     4.1 $& $ 103\pm   54$  \\
  \hline
  \end{tabular}
  \end{center}
  $^a$ in units of $10^{-6}$~s$^{-1}$~cm$^{-2}$\\
  $^b$ unabsorbed, in units of $10^{27}$~erg~s$^{-1}$
  \end{table}

We detect lines of O, Ne, Mg, Si and Fe: The first four elements turn up in the Lyman series and in the He-like triplets; iron can be seen in the ionisation stages \ion{Fe}{xvii} and \ion{Fe}{xviii}. Due to the low signal only five lines can be detected in the HEG and all fluxes agree with the MEG measurement within the statistical uncertainty. In the following we use only the MEG data.

A special analysis is required for the \ion{O}{vii} He-like triplet, because its signal is very weak due to absorption and \emph{Chandra}'s low effective area in this region. Its flux is important to compare IM~Lup to other CTTS and MS stars. There are six photons recorded between 21.5~\AA{} and 22.2~\AA{} (at 21.60, 21.72, 21.80, 22.06, 22.07 and 22.10~\AA{}), which is compatible with the background of 13~cts~\AA$^{-1}$ determined from the line-free region from 20~\AA{} to 23~\AA{} around the \ion{O}{vii} He-like triplet. However, for the line profile of \emph{Chandra} 88\% of all photons are expected within one FWHM=0.02~\AA{}. Three of the six photons fall within this range of the theoretical wavelengths (21.6020, 21.8071 and 22.1012~\AA{} according to the CHIANTI database), but only 0.25 photons are expected from the background in the central FWHM of each line. According to Poisson statistics 3 photons on a background of 0.75 photons are a detection of \ion{O}{vii} emission at the 95\% significance level, but the uncertainty on the derived flux is obviously very large. The 90\% upper limit on the flux, i.e. the flux, where the chance is 10\% to observe three or less photons in the \ion{O}{vii} triplet is $2\times10^{-5}$~cts~s$^{-1}$~cm$^{-2}$, which corresponds to a dereddened luminosity of  $3\times 10^{29}$~erg~s$^{-1}$.

\subsubsection{He-like triplets}
He-like ions show a triplet of lines, where the line ratios are temperature and density sensitive. These triplets consist of a resonance ($r$), an intercombination ($i$) and a forbidden line ($f$) \citep{1969MNRAS.145..241G,2001A&A...376.1113P}. Commonly the $R$- and $G$-ratios ($R = f/i$ and $G = (f+i)/r$) are used; for high electron densities $n_{\mathrm{e}}$ or strong UV photon fields the $R$-ratio falls below its low-density limit, because electrons are collisionally or radiatively excited from the upper level of the $f$ to the $i$ line, but the UV field of late-type stars like IM~Lup is far too weak to influence the $R$-ratio. We interpret the ratios with the CHIANTI~5.1 database.

According to table~\ref{tab1} the observed $R$-ratio for \ion{Si}{xiii} is 1.8, with a 90~\% lower limit of 1.0 according to a Monte-Carlo simulation of Poisson-distributed counts using the observed count numbers as expectation values \citep[for details see][appendix A]{HD163296}. The predicted low-density limit for this ratio is 2.0 with smaller numbers for densities $\log n_{\mathrm{e}} < 12.5$. Not surprisingly the \ion{Si}{xiii} triplet is compatible with the low density limit (90~\% range: $\log n_{\mathrm{e}} < 13.8$).  The G-ratio is temperature sensitive. For temperatures, where more than 10\% of the Si is present in the form of \ion{Si}{xiii}, its value is predicted between 1.2 ($\log T=6.1$) and 0.6 ($\log T=7.2$). This all falls in the $1\sigma$ error range of the observations, thus the $G$-ratio does not constrain the models.

In the \ion{Ne}{ix} triplet the formal $f/i$-ratio is 3, the Monte-Carlo simulation gives lower boundaries on the ratio of 2 ($1\sigma$) and 0.8 (90\%). According to CHIANTI the f/i ratio in the low density regime is 2.8, thus the respective limits on the electron density in the emitting region are $\log n_{\mathrm{e}} < 11.4$ ($1\sigma$) and $\log n_{\mathrm{e}} < 12.3$ (90\%). The intercombination line in the \ion{Ne}{ix} triplet is known to be contaminated with Fe lines in many cases, however, iron is under abundant in IM~Lup and this would further increase the $f/i$ ratio.  The \ion{Ne}{ix} $r$ line also contains a blend from \ion{Fe}{xix}. We cannot correct for this blend because no single \ion{Fe}{xix} is detected and, given the low abundance of Fe, we estimate that this blend is far smaller than the observational uncertainty of the line flux.

In summary, the R-ratio of the \ion{Si}{xiii} triplet limits the electron density of the emitting plasma to $n_{\mathrm{e}} < 13.8$, the R-ratio of the \ion{Ne}{ix} triplet to $\log n_{\mathrm{e}} < 12.3$ (90\% confidence in both cases).

\subsection{The optical spectra}\label{opticalspectra}
The optical spectra are dominated by photospheric absorption lines. From the high-resolution HARPS data we measure the rotational broadening to $v\sin i = 15 \pm 2$~km~s$^{-1}$ by comparison to synthetic PHOENIX spectra \citep{1999JCoAM.109...41H}. Using the stellar radius and the inclination determined by \citet{2008A&A...489..633P} this suggests a stellar rotation period of 8~days if the disk inclination and the stellar inclination match.
We compared the spectra to two templates taken from MS stars of similar spectral type (HD~156274 and HD~156026) in the library of \citet{2003Msngr.114...10B}. The spectra are broadened and degraded to the spectral resolution of our IM~Lup data. The depth of most photospheric lines matches well, thus no significant veiling is present in  IM~Lup. 

\begin{figure}
\resizebox{\hsize}{!}{\includegraphics{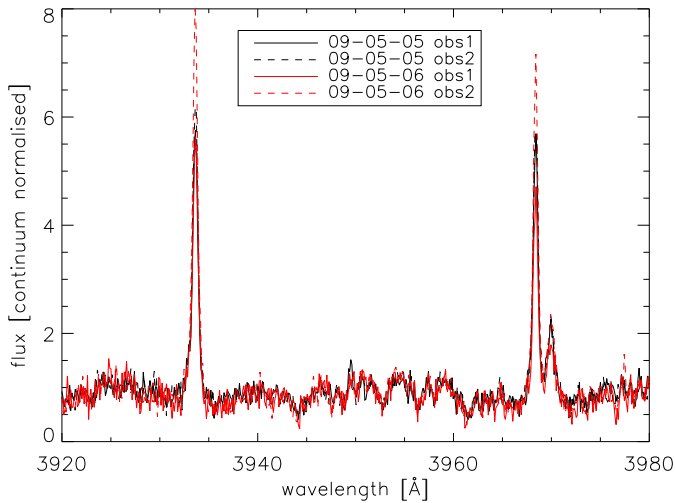}}
\caption{Ca H and K lines in all four exposures. For both lines only sharp emission peaks are visible. The small line at 3970~\AA{} is H$\epsilon$.}
\label{fig_CaHK}
\end{figure}

\begin{figure}
\resizebox{\hsize}{!}{\includegraphics{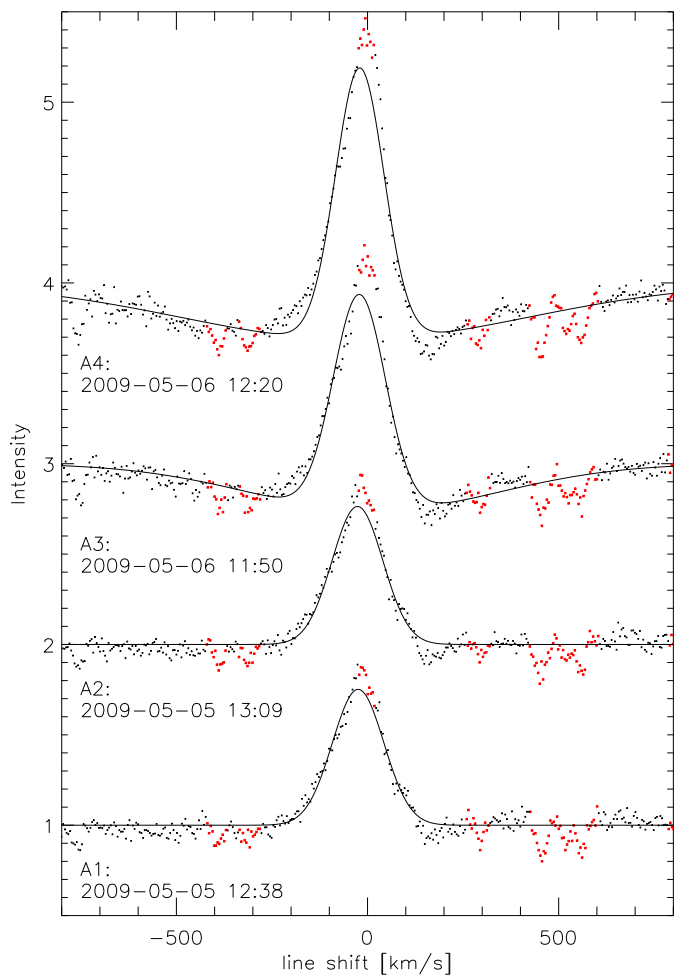}}
\caption{H$\alpha$ line profile in all four exposures. The line is in emission and asymmetric. A broad absorption component is also visible for the second day. Observations are continuum normalised and shifted in intensity for clarity. The red/grey data points are excluded from the fit. Solid lines show a fit.}
\label{fig_ha}
\end{figure}

In the following we discuss where IM~Lup differs from the templates. We find \ion{Li}{i} in absorption at 6708~\AA{}, confirming its youth. The equivalent width of this line is about 0.3~\AA{}. Typical activity tracers like Ca~K and the Balmer series are in emission. Figure~\ref{fig_CaHK} shows the Ca~H and K lines and Fig.~\ref{fig_ha} the H$\alpha$ line profile in the four different exposures taken simultaneously to the \emph{Chandra} observation. 
The H$\alpha$ equivalent width increases during our observations from 2.8~\AA{} on the first day (H1 and H2) to 4.5 (H3) and 5.2~\AA{} (H4) in the exposures on the second day, respectively, where we measured the emission core of the line only (in accordance with the X-ray data we use the convention to denote emission lines with a positive EW and absorption lines with negative values). 

\begin{figure}
\resizebox{\hsize}{!}{\includegraphics{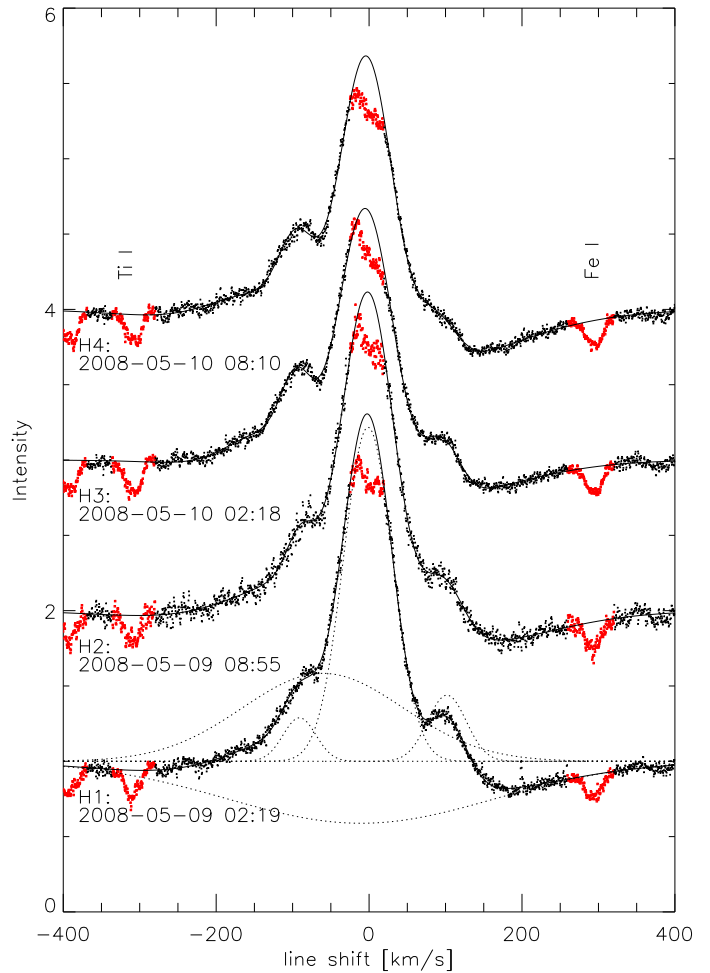}}
\caption{H$\alpha$ line profile in four HARPS exposures one year before our observations. Observations are continuum normalised and shifted in intensity for clarity. The red/grey data points are excluded from the fit. The lines show the best fit of five Gaussians according to table~\ref{tabha}. The single components are indicated for the first observation. Note the smaller velocity scale compared to Fig.~\ref{fig_ha}.}
\label{ha_harps}
\end{figure}

In all cases the H$\alpha$ EW stays below 10~\AA{}, which is the canonical dividing line between CTTS and WTTS. The line profile is asymmetric, with additional emission on the blue side. On the first day a small bump exists at +100~km~s$^{-1}$, which might be due to an unresolved secondary peak at this velocity. This seems to be the typical line profile for IM~Lup; it was also observed by \citet{1999MNRAS.307..909W}. In a HARPS observation (Fig.~\ref{ha_harps}) secondary peaks on both sides of the H$\alpha$ line appear.
On the second day of our ANU observations (A3 and A4) broad absorption wings are present, which are only marginally visible on the first day (A1 and A2), while at the same time the central emission is stronger peaked. 
According to the classification scheme of \citet{1996A&AS..120..229R} the H$\alpha$ line profile in 1998 is III-R \citep{1999MNRAS.307..909W}, in 2008 III-B/R, in 2009 III-R on the first day and IV-B/R on the second day.

In parallel to H$\alpha$ the other detected Balmer lines and also \ion{He}{i} 5867~\AA{} increase in strength on the second day in contrast to the X-ray emission. Surprisingly, Fig.~\ref{fig_lc} shows that the X-ray count rate during the observations on the second day is only half of that on the first day; the hardness ratio is constant. 

For comparison we analyse a HARPS dataset of four exposures of IM Lup, which was taken one year before the \emph{Chandra} observations. The H$\alpha$ emission line, shown in Fig.~\ref{ha_harps}, looks similar; its total equivalent width (including the absorption component) decreases during the observations from 4.3~\AA{} to 2.7~\AA{}. The spectral resolution in these observations is better, thus we can fit the separate components of the line reliably. We exclude two narrow absorption lines from the profile fitting (\ion{Ti}{i} and \ion{Fe}{i}) and also ignore the center of the H$\alpha$ line, which is likely affected by the optical depth. We find that in total five Gaussian components are necessary to describe the line profile: The most prominent component is the central emission peak, then there are two narrow emission components shifted by about 100~km~s$^{-1}$ to the blue and red side, respectively, and two broad components, one of them in absorption. The components are listed in table~\ref{tabha} and shown in Fig.~\ref{ha_harps}. Absorption below the continuum is only seen for velocities above 140~km~s$^{-1}$. Still, the wings on the red side are best fit by a wide absorption component with a centroid close to the stellar rest velocity, if we prescribe a Gaussian shape. If the absorption on the red side in non-Gaussian, than the wide emission component is no longer needed and there is no absorption on the blue side of the line, so that the three narrow emission components are sufficient to describe the line shape. The central emission matches the stellar rest velocity at all times, its amplitude changes with time. In the fit this is seen as decreasing flux in the wide component, while the central component is nearly constant. The narrow red-shifted component moves towards the center during the observations, while the narrow, blue-shifted emission keeps its position.

In the ANU observations a qualitatively similar structure is observed, but the resolution and signal-to-noise ratio is lower, so only two Gaussians are required. Absorption seen on both sides of the line on the second day of the ANU observations (A3 and A4 in Fig.~\ref{fig_ha}) is consistent with a wide absorption component centred on the stellar rest velocity and sigificantly broader than in the HARPS dataset. The FWHM of the main emission feature of 160~km~s$^{-1}$ and the velocity shift of the bump on the red side of the line profile in the first two observations agrees with the narrow red component in the HARPS observations (table~\ref{tabha}). The absorption in the ANU observations is stronger and broader than in the HARPS data, in fact, it is stronger than the emission line on the second day. Gaussian profiles are not ideal to fit the line in Fig.~\ref{fig_ha}, the fit is too shallow and too broad on the red side around 150~km~s$^{-1}$, but they illustrate the change in relative strength between the observations. The fits underpredict the flux in the line center, because the line is more strongly peaked than a Gaussian.

\begin{table} [h]
\caption{Properties of line profile components \label{tabha}}
\begin{center}
\begin{tabular}{llrrr}
\hline \hline
Comp. & No & EW$^a$ & centre & FWHM \\

& & [\AA] & [km s$^{-1}$] & [km s$^{-1}$] \\
\hline
\multicolumn{5}{c}{HARPS - fit with 5 Gaussians}\\
\hline
central               & H1 & $    3.9 \pm    0.1 $ & $  -2.6 \pm    0.2 $ & $   76 \pm   1 $\\
                      & H2 & $    3.6 \pm    0.2 $ & $  -1.7 \pm    0.4 $ & $   77 \pm   2 $\\
                      & H3 & $    3.9 \pm    0.1 $ & $  -5.7 \pm    0.2 $ & $   85 \pm   1 $\\
                      & H4 & $    3.9 \pm    0.1 $ & $  -4.2 \pm    0.3 $ & $   82 \pm   1 $\\
blue wide             & H1 & $    3.4 \pm    1.1 $ & $  -62 \pm     9 $ & $  248 \pm  13 $\\
                      & H2 & $    4.4 \pm    3.8 $ & $  -53 \pm    22 $ & $  249 \pm  30 $\\
                      & H3 & $    0.8 \pm    0.1 $ & $ -133 \pm     3 $ & $  115 \pm   4 $\\
                      & H4 & $    1.5 \pm    0.4 $ & $ -128 \pm     7 $ & $  166 \pm  10 $\\
blue narrow           & H1 & $   0.34 \pm   0.02 $ & $ -92.2 \pm    0.6 $ & $   51 \pm   2 $\\
                      & H2 & $   0.29 \pm   0.04 $ & $ -90.5 \pm    1.2 $ & $   43 \pm   4 $\\
                      & H3 & $   0.55 \pm   0.02 $ & $ -95.6 \pm    0.4 $ & $   49 \pm   1 $\\
                      & H4 & $   0.62 \pm   0.04 $ & $ -95.5 \pm    0.5 $ & $   54 \pm   2 $\\
wide abs              & H1 & $   -3.9 \pm    2.3 $ & $  -12 \pm     4 $ & $  412 \pm  24 $\\
                      & H2 & $   -4.6 \pm    5.3 $ & $   -5 \pm    16 $ & $  354 \pm  43 $\\
                      & H3 & $   -2.2 \pm    0.3 $ & $   36 \pm     3 $ & $  312 \pm   6 $\\
                      & H4 & $   -3.8 \pm    0.7 $ & $    7 \pm     4 $ & $  351 \pm   7 $\\
red narrow            & H1 & $   0.61 \pm   0.04 $ & $ 100.6 \pm    0.5 $ & $   60 \pm   2 $\\
                      & H2 & $   0.51 \pm   0.09 $ & $  96.3 \pm    1.4 $ & $   60 \pm   5 $\\
                      & H3 & $   0.48 \pm   0.01 $ & $  93.8 \pm    0.4 $ & $   55 \pm   1 $\\
                      & H4 & $   0.47 \pm   0.03 $ & $  86.2 \pm    0.9 $ & $   62 \pm   2 $\\
\hline
\multicolumn{5}{c}{ANU - fit with 1 or 2 Gaussians}\\
\hline
central    & A1 & $    2.8 \pm    0.2 $ & $ -25.8 \pm    1.0 $ & $  159. \pm     3. $\\
           & A2 & $    2.8 \pm    0.2 $ & $ -27.5 \pm    1.0 $ & $  157. \pm     2. $\\
           & A3 & $    4.5 \pm    0.2 $ & $ -22.6 \pm    0.6 $ & $  161. \pm     2. $\\
           & A4 & $    5.2 \pm    0.2 $ & $ -21.7 \pm    0.5 $ & $  150. \pm     1. $\\
absorption & A3 & $   -4.7 \pm    2.8 $ & $   15. \pm     7. $ & $  750. \pm    50. $\\
           & A4 & $   -7.8 \pm    5.1 $ & $  -36. \pm     6. $ & $ 1050. \pm    50. $\\
\hline
\end{tabular}
\end{center}
$^a$ absorption is shown with negative values
\end{table}

\subsection{Comparison X-ray properties to CTTS, WTTS and MS stars}
To compare the X-ray luminosity with other TTS of similar spectral type we used the data published from the COUP \citep{2005ApJS..160..401P} and the XEST \citep{XEST} projects. Both are large X-ray surveys covering the star forming regions of Orion and the Taurus molecular cloud, respectively. IM~Lup is a bright object compared to both datasets.

Compared to the Taurus molecular cloud \citep{2007A&A...468..425T} the mean temperature and X-ray luminosity $L_X$ of IM~Lup would make it the most luminous CTTS. If it is a WTTS it still belongs to the brightest 20\% of all WTTS in that region. 

The bolometric luminosity of  IM~Lup is $L_{bol}=1.9 L_{\sun}$ \citep{2008A&A...489..633P} so $\log (L_X/L_{bol})=-3.1$. This number is typical both for MS stars, which saturate at $\log (L_X/L_{bol})=-3$, and also for CTTS or WTTS \citep{2003A&A...402..277F,2005ApJS..160..401P}.

The abundances of IM~Lup shown in table~\ref{tabfit} exhibit an IFIP (inverse first ionisation potential) pattern, where elements of large first ionisation potential are enriched relative to the solar abundance and those of low FIP are depleted. This is observed in most active stars, independent of their evolutionary state. The values found in IM~Lup agree with typical abundances as given in the review by \citet{2004A&ARv..12...71G}.

\citet{RULup} and \citet{manuelnh} present an analysis of the \ion{O}{viii}/\ion{O}{vii} ratio as a measure of the excess of soft emission in CTTS with respect to stars on the MS. These studies show the observed excess to be confined to a narrow temperature range at the formation of the He-like \ion{O}{vii} triplet at 1-2~MK. 
The upper limit we determine for the \ion{O}{vii} emission in IM~Lup is compatible with both,  MS stars and the soft excess in CTTS.

The $f/i$ ratio of the \ion{Ne}{ix} triplet in IM~Lup is statistically compatible on the $1\sigma$ level with typical values of coronal sources \citep{2004A&A...427..667N} in the low-density limit; however, a reduced $f/i$ value is allowed for IM~Lup on the 90\% confidence level due to the low count statistic. The low-density limit is also observed in WTTS (TWA~4: \citep{2004ApJ...605L..49K};  TWA~5: \citep{twa5} and \object{HDE 245059} \citep{2009ApJ...697..493B}). Most CTTS on the other hand show a reduced $f/i$ ratio which is attributed to their accretion spots on the surface. TW~Hya \citep{2002ApJ...567..434K,twhya}, BP~Tau \citep{bptau}, V4046~Sge \citep{v4046}, RU~Lup \citep{RULup}, MP~Mus \citep{2007A&A...465L...5A} and Hen~3-600 \citep{2007ApJ...671..592H} all show significant deviations from the low-density limit. 

Going beyond the He-like triplets we compared our MEG spectrum (Fig.~\ref{fig_meg}) of IM~Lup with a \emph{Chandra}/MEG spectrum of \object{YY Gem} \citep{2008A&A...491..859L}, a MS binary of spectral type M3-M4, similar to IM~Lup. After imposing the absorbing column density found for IM~Lup both spectra show the same features. The remaining small differences in the line fluxes are due to small abundance effects and IM~Lup is a little stronger in the short-wavelength continuum, because of its strong, hot corona. Notable differences turn up, when the same experiment is done with the spectrum of TW~Hya, because it lacks the strong emission component at 2.1~keV.

\section{Discussion}
\label{discussion}

\subsection{X-ray properties}
IM~Lup shows many characteristics of an active star. It has a bright, hot corona and its abundances follow the IFIP pattern. These traits are common to both CTTS and WTTS. In the density diagnostic of the He-like triplets there is no deviation from the coronal limit, however, due to the low signal a high-density state is allowed at the 90\% confidence level. The signal of the \ion{O}{vii} triplet is too weak to check for any soft X-ray excess. 
Thus, we conclude that -- from the X-ray point-of-view -- it is more likely that IM~Lup shares its characteristics with WTTS, despite the presence of its disk. If confirmed, this suggests that the distinctive characteristics of CTTS are really due to accretion: An excess of soft X-ray emission originates in the high-density environment of an accretion spot as simulated by \citet{lamzin} and \citet{acc_model}. Our observation is incompatible with any hypothetical scenario where the high-density signatures of the He-like triplets are attributed to the accretion disk itself because they are not observed (on the 1$\sigma$ confidence level) in IM Lup, which has a disk.

The only peculiarity in comparison to the WTTS samples of COUP and XEST is the high bolometric luminosity of IM~Lup. It belongs to the brightest objects of its kind. This may be due to an overestimation of the distance. We use $190\pm27$~pc from \citet{1998MNRAS.301L..39W} based on the \emph{HIPPARCOS} parallax, but other methods lead to lower values of $140\pm20$~pc \citep{1993AJ....105..571H}. Both values are compatible within the $2\sigma$ errors, but if the lower value was the true one, X-ray emission measure and bolometric luminosity would be only half of the values given above. This does not change any of the other conclusions of our analysis, but IM~Lup would fit in better with the bulk of the COUP and XEST objects in this case. Still, IM~Lup is an active star with a bright and hot corona and $\log (L_X/L_{bol})=-3.1$, similar to saturated MS stars and also typical for CTTS \citep{2009A&ARv..17..309G}. 

\subsection{Mass accretion?}
\label{massacretion}
IM~Lup has a high level of chromospheric activity, which causes the emission observed in H$\alpha$. Other M dwarfs with active chromosphere also show H$\alpha$ emission, typically with an equivalent width of a few~\AA{} \citep{1981ApJS...46..159W, 1990ApJS...74..891R}, very similar to IM~Lup. Their Ca~H and K lines often show a very narrow emission peak, too. The H$\alpha$ line profiles vary, but in  most active stars there is a single, narrow emission peak, sometimes with a reversal due to self-absorption in the centre. This is in line with the central peak in the IM~Lup data. In the samples of \citet{1981ApJS...46..159W} and \citet{1990ApJS...74..891R} there is not a single case of H$\alpha$ in emission, where simultaneously broad absorption is present. The same is true for WTTS, which have narrow emission lines of symmetric shape \citep{2009A&A...504..461F}.

In contrast, red-shifted absorption dips and red- and blue-shifted emission components appear in accretion models in most viewing geometries \citep{2001ApJ...550..944M}. Thus, in IM~Lup the H$\alpha$ line profile must be caused by additional material which is not present in MS stars: Candidates are a circumstellar envelope, the disk or accretion funnels. We do not detect veiling of the photospheric emission lines, thus any mass accretion must be very low, if present at all. Still, the narrow emission components might originate in  a weak accretion funnel. They are shifted by $\pm100$~km~s$^{-1}$, because we see accretion towards both, the observers side of the star and the back side. The line shift is smaller than usually seen in emission lines from accretion funnels of CTTS, but given the uncertain stellar parameters of IM~Lup this could be due to a combination of low mass, i.e. smaller gravitational potential, and viewing geometry, where the accretion funnel is not parallel to the line-of-sight. 

In some stars the H$\alpha$ line profile is also influenced by strong flares but this cannot be the case in IM~Lup, because no flare shows up in the X-ray lightcurve.

The study of \citet{2003ApJ...582.1109W} correlated the width of the H$\alpha$ line at 10~\% of the maximum with accretion. They suggest a full width at 10~\% of the maximum $>270$~km~s$^{-1}$ as borderline between accreting and non-accreting sources, a slightly lower limit of $>200$~km~s$^{-1}$ was presented by \citet{2003ApJ...592..282J}. The line width of IM~Lup falls just between those two limits; it would be classified as non-accreting by the one and as accreting by the other boundary.
\citet{2004A&A...424..603N} showed that a quantitative relation between line width and mass accretion rate exists, which is valid from CTTS to sub-stellar objects. This relation yields an accretion rate of $10^{-11} M_{\sun}\;\mathrm{yr}^{-1}$, but IM~Lup is chromospherically very active, which explains a large fraction of the H$\alpha$ flux. Additionally, not all relations employed to convert the H$\alpha$ flux to a mass accretion rate agree, if accretion is present at all, therefore we treat this number as a rough estimate only.

In summary, the H$\alpha$ EW classifies IM~Lup as a WTTS, but the H$\alpha$ line profile is complex as in CTTS. This suggests that mass accretion occurs at a very small rate. The energy released from accretion is too low to influence the X-ray properties significantly. For a mass accretion rate of $10^{-11} M_{\sun}\;\mathrm{yr}^{-1}$, at most 0.5\% of the X-rays can be due to accretion, assuming that 10\% of the accretion energy is released in X-rays \citep{acc_model}. For the interpretation of the X-ray data we can thus safely treat IM~Lup as a non-accreting source. 

\subsection{Varying absorption?}
The HARPS H$\alpha$ line profiles are presented in Fig.~\ref{ha_harps}. While the fitting with Gaussians (table~\ref{tabha}) yields an absorption component reaching over the entire line profile, for non-Gaussian components absorption is only required on the red side where the observed line profile falls below the continuum. The situation is different in the ANU data, which was obtained simultaneous to the \emph{Chandra} observation. No absorption is visible here on the first day and wide absorbtion wings are present on both sides of the line on the second day. 
We postulate that matter from the disk on its way onto the star causes the H$\alpha$ absorption. With IM~Lup's mass and radius (Sect.~\ref{imlupprop}) we calculate the Keplerian rotation of the disk.
At the co-rotation radius of $6\;R_{*}$ it matches the stellar rotation period from Sect.~\ref{opticalspectra}; in this region the inner disk should be truncated. The absorption component is strong at 100-400~km~s$^{-1}$, which corresponds to the free-fall velocity from the co-rotation radius. Thus the H$\alpha$ absorbing gas may be found in the gap between star and disk; however, this scenario can only explain the red-shifted part of the absorption, but it strengthens the argument in Sect.~\ref{massacretion} that the H$\alpha$ line width at 10\% maximum could be interpreted as accretion. We speculate, that the blue-shifted absorption in the ANU data is caused by an outflow. CTTS commonly drive winds or even jets, where the mass loss is of the order of 10\% of the accretion rate \citep{1990ApJ...354..687C,2008ApJ...689.1112C}. The blue-shifted absorption appears only on the second day of the ANU observation, when the absoprtion on the red side, which we interpret as accretion, is stronger. 

We divided the \emph{Chandra} observation in two parts and fitted the $n_H$ indenpedently to test if the absorption in the X-ray data changes with time. We found no change of the fitted value. Also, the hardness ratio of the X-ray spectra does not change during the observations. This does not rule out variability in $n_H$, but is sets an upper limit. We estimate that an additional absorbing column of $>10^{21}$~cm$^{-2}$ would have been detected. Does the accretion scenario explain why strong H$\alpha$ absorption coincides with weak X-ray emission? Given the weak limit on variability in $n_H$ and the fact that we have only two datapoints of simultaneous X-ray and optical spectra, the relation between H$\alpha$ absorption and X-ray luminosity is --at best-- suggestive. Still it is intriguing to interpret the lower X-ray luminosity and the H$\alpha$ line profiles as absoption by an accretion funnel. This should cause a hardening of the X-ray absorption. This is not observed, but the limit on extra absorption is weak due to the low count rates.

If taken seriously, the relation between X-ray luminosity and H$\alpha$ absorption fits to the fact that  
CTTS are on average less luminous in X-rays than WTTS \citep[e.g.][]{2005ApJS..160..401P,2007A&A...468..443T}.  \citet{2009ApJ...699L..35D} suggested that it is not the accreting nature of the CTTS that causes their low luminosity; rather, it is the low luminosity that allows the disk to reach closer to the star, i.e., weaker emitters accrete and become CTTS.

\subsection{IM Lup in context}
\citet{2009A&A...501.1013S} observed variability in optical spectra of \object{T Cha}. Like IM~Lup T~Cha is also a transition object from CTTS to WTTS without veiling of photospheric lines, i.e. with little or no mass accretion. \citet{2009A&A...501.1013S} find the H$\alpha$ profiles to change from emission lines to absorption components over time, simultaneously they observe changes in the reddening and develop a scenario based on inhomogeneous circumstellar extinction similar to our ideas for IM~Lup. They argue this to be caused by the inner parts of the disk, where grain growth takes place and possibly planets form. 

Only a small number of CTTS have been observed with X-ray grating spectroscopy, most of them show low $f/i$ ratios in the He-like triplets, but \object{T Tau} does not. Does IM~Lup confirm  a CTTS population with low densities like T~Tau or does it behave as WTTS? Although the complex H$\alpha$ line profile points to mass accretion, IM~Lup differs from T~Tau in other characteristics. The main component of the multiple system T~Tau is much more massive, close to the boundary to the HerbigAe/Be stars and it accretes at much higher rates. Whereas IM~Lup's X-ray emission must be coronal, T~Tau could power a significant part of its luminosity from accretion.  We therefore expect IM~Lup to resemble the WTTS and young MS stars in its X-ray properties. Again, only very few WTTS are observed with X-ray gratings, but they all look similar to MS stars. Despite the small number, it is credible that their emission is driven exclusively by a solar-type corona and we find no evidence that the transition object IM~Lup, which still posseses a disk, is different. The transition from a CTTS to a WTTS in X-rays seems to coincide with the decline of accretion as expected from accretion shock models for CTTS. However, the \ion{Ne}{ix} triplet is the strongest discriminator and the original driver of the reported observation, but at the 90\% confidence level the CTTS-like high-density scenario cannot be excluded for IM~Lup.

\section{Summary}
\label{summary}
We present a deep \emph{Chandra}/HETGS grating spectrum of IM~Lup, a WTTS with a circumstellar disk, and accompanying optical spectroscopy, where some of the spectra were obtained simultaneous to the X-ray spectra. IM~Lup is an active star with an IFIP effect in a hot corona. The \ion{Ne}{ix} He-like triplet is in the low-density limit, but due to the low signal we also cannot exclude a low $f/i$ scenario. It seems, that the X-ray signatures of CTTS, a soft excess and high-density He-like triplets, are not correlated with the presence of the disk, but with the active accretion of CTTS. IM~Lup's optical spectra exhibit emission lines, which are typical for chromospherically active stars. This fits well with the high X-ray luminosity of IM~Lup. However, the H$\alpha$ line profiles are more complex than expected in MS chromospheres. This could possibly be due to weak accretion funnels as in CTTS with a low accretion rate.	

\begin{acknowledgements}
CHIANTI is a collaborative project involving the NRL (USA), RAL (UK), MSSL (UK), the Universities of Florence (Italy) and Cambridge (UK), and George Mason University (USA). We made use of observations made with ESO Telescopes at the La Silla Observatory under programme ID 081.C-0779(A). The authors would like to thank Ian Waite, USQ, for the initial extraction of the IM Lup data from ANU. H.M.G. acknowledges support from DLR under 50OR0105. SPM was supported by an appointment to the NASA Postdoctoral Program at Ames Research Center, administered by Oak Ridge Associated Universities through a contract with NASA. Z.-Y.L. acknowledges support from Chandra grant GO9-0007X.
\end{acknowledgements}

\bibliographystyle{aa} 
\bibliography{../articles}
\end{document}